\documentclass[namedreferences,noid]{SolarPhysicsarx}
\usepackage{graphicx}        
\usepackage{color}           
\usepackage{url}             

\begin{document}

\begin{article}

\begin{opening}

\title{Distribution of the daily Sunspot Number variation for the last 14 solar cycles}

\author{Mihail-Ioan~\surname{Pop}$^{1}$}

   \institute{$^{1}$ Department of Physics, Transilvania University of Brasov,
                     email: \url{mihailp@unitbv.ro}
             }

\begin{abstract}
The difference between consecutive daily Sunspot Numbers was analysed. Its distribution was approximated on a large time scale with an exponential law. In order to verify this approximation a Maximum Entropy distribution was generated by a modified version of the Simulated Annealing algorithm. The exponential approximation holds for the generated distribution too. The exponential law is characteristic for time scales covering whole cycles and it is mostly a characteristic of the Sunspot Number fluctuations and not of its average variation.
\end{abstract}
\end{opening}
\section{Introduction}
\begin{figure}
 \centerline{ \includegraphics[width=0.95\textwidth,clip=]{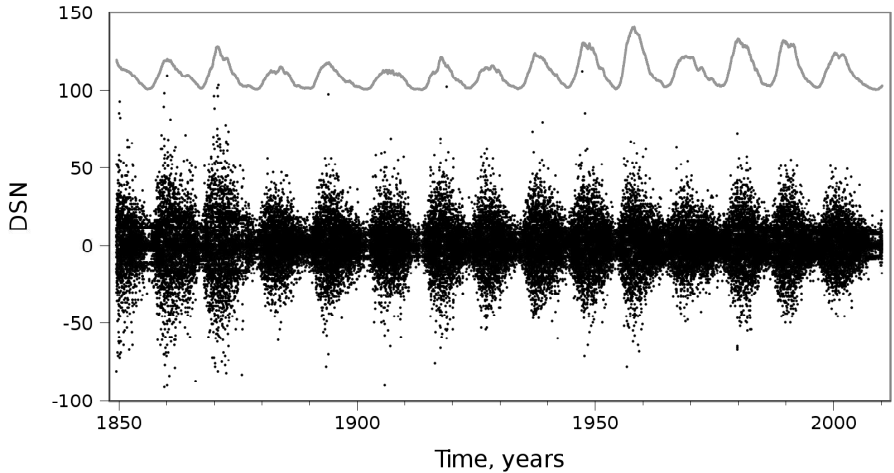} }
 \caption{Evolution of $DSN$ (black points) along solar cycles (average $SN$ - grey line, not to scale). The graph starts in the middle of the $12^{th}$ cycle and spans for the last 14 cycles.}
 \label{DSN}
 \end{figure}

The oldest and most used index for solar activity is the Sunspot Number ($SN$) or Wolf number describing the number of sunspots $S$ and groups of sunspots $G$ that can be observed on the Sun: $SN = 10 G + S$. The $SN$ index is computed daily and there is a complete record of values $SN_i$ since the 1850's, spanning Solar Cycles 10-23. For the following we used the difference between daily $SN$ numbers:
\begin{equation}
DSN _i = SN _i - SN _{i-1},
\end{equation}
where $i$ is the count of days. The $SN$ values have been obtained from the Solar Influences Data Analysis Center of the Royal Observatory of Belgium \cite{SIDC}. The graph of $DSN$ as a function of time is presented in Figure \ref{DSN}. The distribution of $DSN$ is linked to the mechanisms that generate sunspots with a periodicity of about 11 years. While there are deterministic mechanisms involved, there are also some stochastic mechanisms, responsible especially for the high day to day variability of $SN$. Thus, a statistical analysis of $DSN$ may reveal some characteristics of the latter type of mechanism. We will further refer to $DSN$ values and their characteristics determined from measured $SN$ values as "empirical" values or characteristics.

The distribution of $DSN$ has been reconstructed both as a histogram and as a Maximum Entropy Method (MEM) distribution (\opencite{Jaynes1},\citeyear{Jaynes2,Jaynes}; \opencite{Uffink}). The MEM distribution was built by imposing some local conditions and was determined using a modified form of the Simulated Annealing (SA) heuristic search algorithm \cite{Kirk,Cerny,Ingber}. The details of the MEM distribution are presented in Appendix A.

The values of $DSN$ as computed from $SN$ values are integer numbers. However, we argue that both $SN$ and $DSN$ values can be thought as real numbers, where the fractional part expresses the degree with which a certain "average sunspot" is realised. Thus, the maximum entropy distribution was extended into the set of real numbers. The distributions obtained were approximated on certain ranges of $DSN$ with exponential functions of the form:
\begin{equation} \label{Exp}
\rho(DSN) = \frac{1}{Z} \exp \left( -\frac{DSN}{m} \right),
\label{Aprox}
\end{equation}
where $m$ is a distribution parameter and $Z$ is a constant such that $\rho$ has integral equal to 1. 

The $SN$ time series have both a deterministic component, which may be taken as the monthly or yearly average of $SN$, and a stochastic component, the latter appearing as daily fluctuations of $SN$ around the average. Both components have characteristic distributions. The basic models used in explaining the long-term (on scales of months and above) behaviour of the solar cycle are the dynamo models \cite{Proctor}. It was pointed out that this model may exhibit chaotic behaviour. \inlinecite{Letellier} discuss such a chaotic mechanism. Their conclusions are that it has low dimensionality and similar behaviour to a R\"ossler system of differential equations. On the other hand, \inlinecite{Aguirre} model the solar cycle, transformed into a symmetrical space, through an autoregressive model with deterministic terms and stochastic residuals with Gaussian distribution. The model obtained has a chaotic attractor similar to that of the R\"ossler system. Aguirre, Letellier, and Maquet point out that the sunspot number time series is nonstationary. \inlinecite{Nordemann} approximated the time series of yealy averages of $SN$ with an exponential function separately for the rising and declining phases. The exponential dependence indicates a power law distribution of the $SN$ yearly averages for each phase. For larger time scales the distribution may deviate from the power law.

The stochastic part of the $SN$ time series has been analysed both in connection to a chaotic dynamo and as a separate term. In many studies exponential distributions and dependences were found. \inlinecite{Pontieri} build a dynamo model with fractional Brownian motion in order to explain the stochastic character of the $SN$ time series. They determine the Hurst exponent of $SN$ to be $H \approx 0.76$ on time scales from 20 to 350 days. This indicates a Brownian motion with positive autocorrelation at work at these time scales. \inlinecite{Noble} develop a model for the stochastic part of the $SN$ time series which is represented as a diffusion process described by the Fokker-Planck equation. The solution of this equation has an approximate exponential form. They find that at solar minima the $SN$ index has a distribution with exponential tail. They also point out that the daily $SN$ distribution is approximately exponential for the period 1850-2010. \inlinecite{Greenkorn} looked at daily $SN$ data in order to determine the type of flow in the Sun's convective layer. The $SN$ values indicate stochastic behaviour for Cycles 10-19 and chaotic behaviour for Cycles 20-23, the last of which presents an increase in the stochastic character compared to previous cycles. 

Salakhutdinova \shortcite{Salakhutdinova1,Salakhutdinova2} analysed the $SN$ time series by separating it into a regular (deterministic) low frequency component and a stochastic high frequency component. The stochastic component has properties of white (Gaussian) noise at time scales lower than two months and for time scales between two months and two years it has properties similar to pink (flicker) noise. The two kinds of properties may characterize two sub-components that combine to form the stochastic part of $SN$. Generally, for large time scales the stochastic component can be associated with a chromatic noise with sharp maxima. The regular component behaves as a nonlinear oscillator, with similarities to a R\"ossler system. \inlinecite{Lepreti} compute the Hurst index for the $SN$ index and the H$\alpha$ flare index for the period 1976-1996. The two indices have similar Hurst values ($0.76$ and $0.74$) on intervals of 20-350 days and 24-450 days respectively. These show the presence of long-range autocorrelations for the two solar activity indices for the period 1976-1996. The authors determined the distribution of normalised fluctuations of the $SN$ and H$\alpha$ flare indices. The fluctuations were computed as 1-day differences of the respective indices. Next, these differences were normalised by subtracting their average value and then dividing by their root mean square value. For both indices, the probability density functions are stretched exponentials of the form $\rho (x) = A \exp \left( - a |x| ^r \right)$. For $SN$ they found $r \approx 1.04$, which is very close to a Laplace distribution and for the flare index $r \approx 0.66$. 

\inlinecite{Vecchio} analyse the coronal emission time series at 530.3 nm from 1949 to 1996 with the Proper Orthogonal Decomposition method. The method uses a series expansion of the coronal emission relative to some basis functions by means of coefficients which depend on time. The authors determined the distribution of the 1-day difference of each coefficient's values. The probability density function is similar to a stretched exponential for the central part of the distribution, where $r$ is close to 1. 

\inlinecite{Kanazir} model the flare energy distribution and waiting times for two solar active regions spanning each about one month. They find that the flare frequency-size index has a power law distribution, while the waiting times have an exponential distribution. The parameter of the exponential distribution may suffer a jump in time, as found for one of the two active regions. This jump is explained in a jump-transition model by corresponding step changes in the rates of energy input and flare transition. Nevertheless, between step changes the waiting times distribution is a stable exponential distribution.

\section{Empirical Data Approximations}
\begin{figure}
 \centerline
 {
 	\includegraphics[width=0.5\textwidth,clip=]{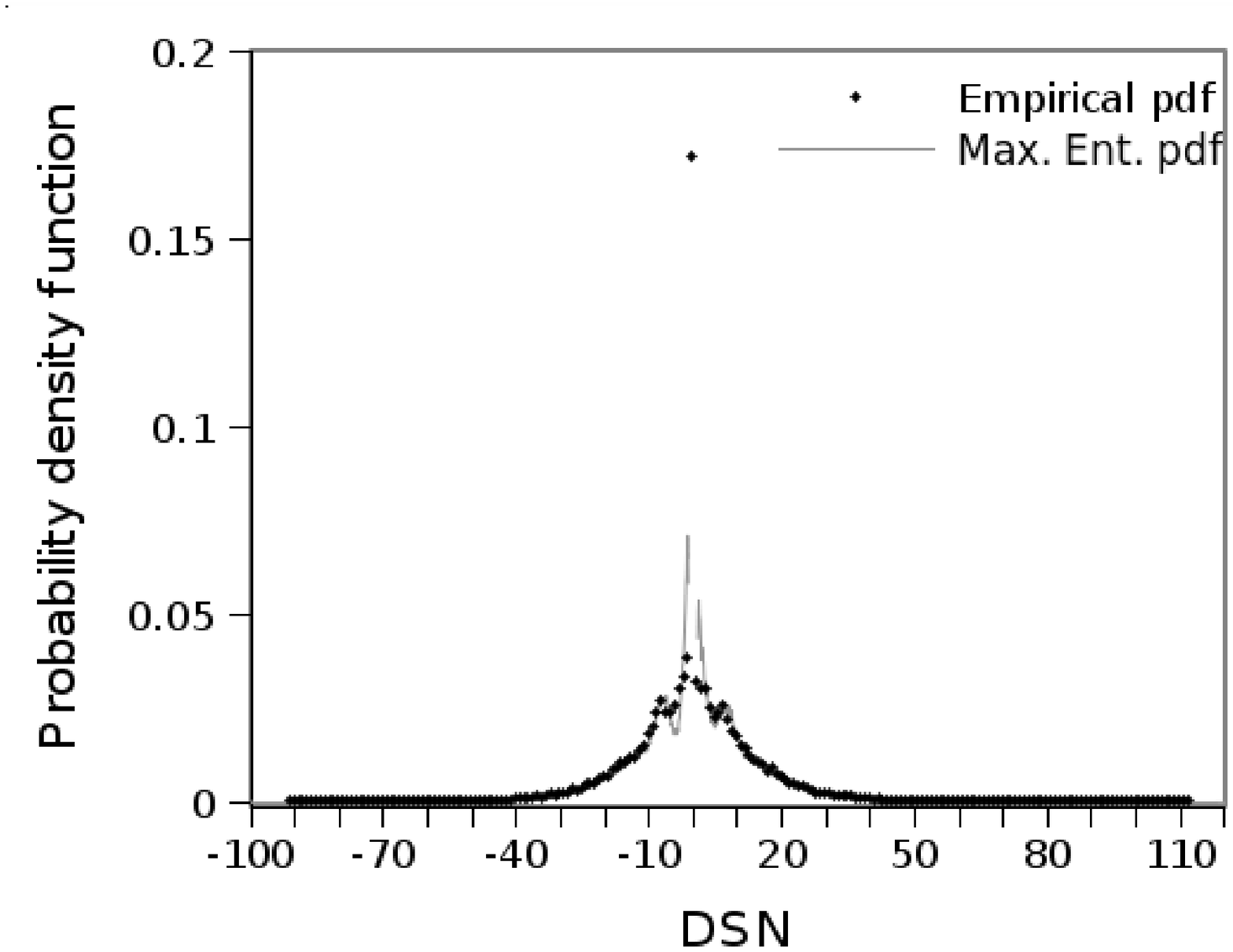}
 	\includegraphics[width=0.5\textwidth,clip=]{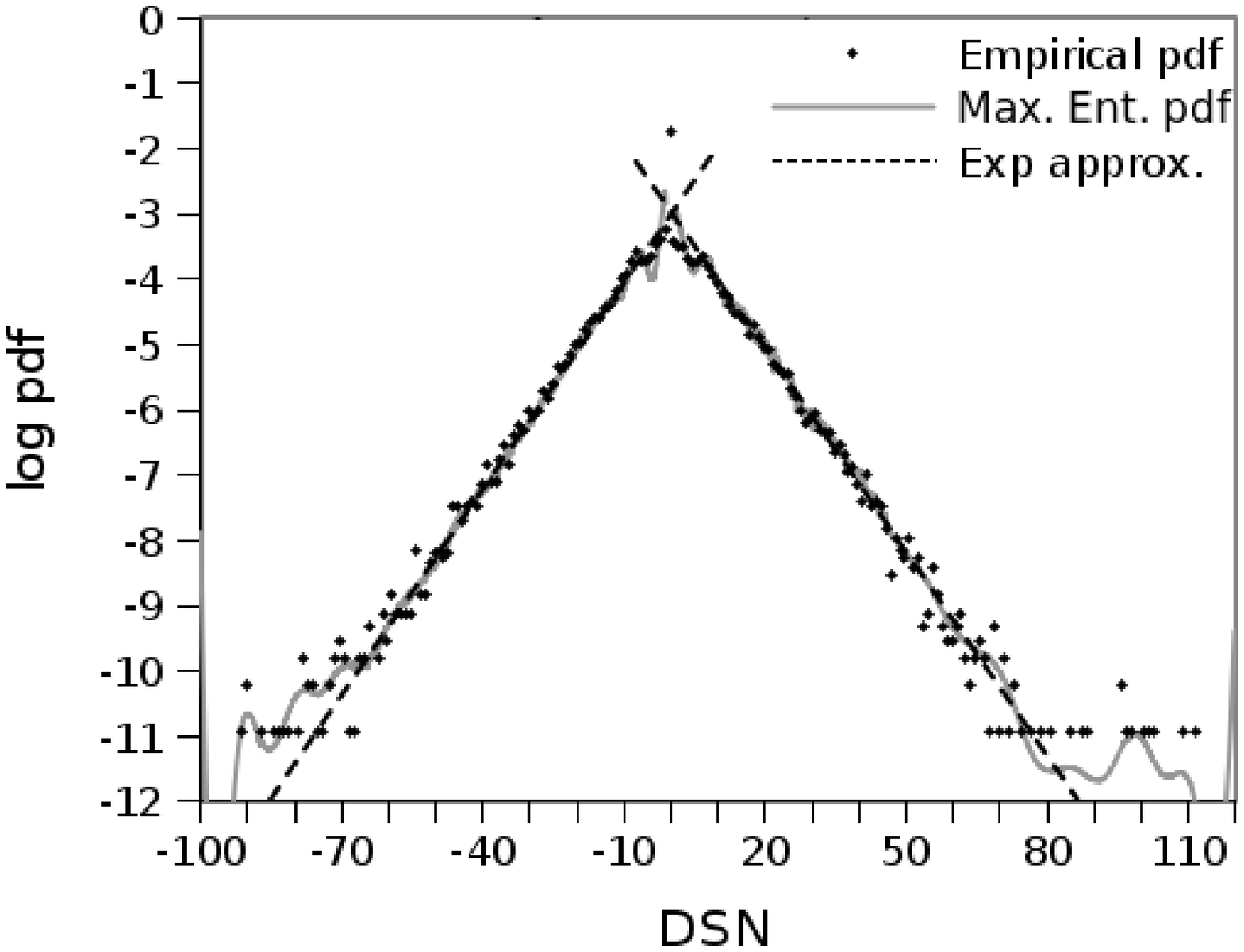}
 }
 \caption{(left) Empirical and Maximum Entropy distributions of the values of $DSN$. Note the high pdf value associated with $DSN = 0$; (right) Natural logarithm of the $DSN$ distributions with exponential approximations of the Maximum Entropy pdf computed for $|DSN| \in [10;60]$.}
 \label{roMEM} \label{lnroMEM}
 \end{figure}

In order to obtain the distribution of $DSN$, daily data from Cycles 10 - 23 were used.  The values of $DSN$ range from -91 to 112. The empirical values of $DSN$ were computed from the registered $SN$ values and then the empirical probability $p(DSN)$ was determined by counting the occurrence of each value of $DSN$. The probability for each $DSN$ is presented in Figure \ref{roMEM}. By applying a least squares fit to $\ln p(DSN)$, it was found that $p(DSN)$ can be approximated well with exponential functions (\ref{Exp}) with the following parameters: $m = - 9.32$ for $DSN \in [-60;-10]$ and $m = 9.37$ for $DSN \in [10;60]$. The goodness of fit is characterised by the correlation coefficient $R \approx 0.99$ in both cases. Outside these intervals $p(DSN)$ diverges from the exponential fit. A special case is $DSN = 0$, for which the probability is much bigger than probabilities associated to other values of $DSN$. Thus, it appears that $DSN = 0$ represents a special state. We propose a model to describe the general behaviour of $DSN$, according to which the probability density function (pdf) of values of $DSN$ in the set of real numbers is of the form:
\begin{eqnarray}
\rho(DSN) &=& p(DSN = 0) \delta(DSN) + p(DSN < 0) \rho^-(DSN) \nonumber \\
           && + p(DSN > 0) \rho^+(DSN),
\end{eqnarray}
where $\delta(\cdot)$ is the Dirac distribution, $\rho^-(\cdot)$, $\rho^+(\cdot)$ are probability density functions associated with negative and positive values of $DSN$ respectively and $p(\cdot)$ are probabilities associated with the respective intervals of $DSN$. From the empirical distribution determined above, $p(DSN = 0) = 0.172$. Since $p(-DSN) \approx p(DSN)$, we take $p(DSN < 0) = p(DSN > 0) = 0.414$. The density functions $\rho^-$, $\rho^+$ have been determined by the Maximum Entropy Method (see Appendix A). Note that the empirical probabilities $p(DSN)$ can be taken as (smoothed) pdf values associated with the corresponding values of $DSN$ since pdf is probability divided by corresponding interval and $p(DSN)$ corresponds to an interval of $DSN$ of length 1.

\section{Results}

\subsection{Maximum Entropy distribution approximations}

We collected $55\,882$ empirical values of $DSN$. In order to determine the pdfs $\rho^-(\cdot)$, $\rho^+(\cdot)$ all values of 0 were eliminated, according to model (3). We used $N^- = 23\,587$ $DSN$ values for $\rho^-(\cdot)$ and $N^+ = 22\,696$ values for $\rho^+(\cdot)$. The pdf estimate obtained is presented in Figure \ref{roMEM}; also, its logarithm is represented along exponential fits computed by the least squares method for $DSN \in [-60;-10]$ and $DSN \in [10;60]$ respectively. The fits have parameters $m = - 9.51$ for $DSN \in [-60;-10]$ and $m = 9.60$ for $DSN \in [10;60]$. The goodness of fit is characterised by $R \approx 0.995$. These are very close to the empirical data fits. We conclude that indeed there is an exponential law in the distribution of $DSN$.

From both the empirical values and the MEM distribution it can be seen that the distribution of $DSN$ changes abruptly from the exponential approximations for small $|DSN|$. There are two local maxima at about $DSN = -7$ and $DSN$ between 7 and 8. There may be an observational cause for this characteristic. When the $SN$ index is computed from sunspot observations, each group of sunspots counts as 10 individual sunspots, thus it may be easier to observe a higher value, e.g. $SN \geq 7$ than smaller values. Thus, jumps from previous values of $SN = 0$ to $SN = 7$ and back may be more probable than other jumps around the cycles' minima.

For values of $|DSN| < 7$ the MEM distribution registers a variation that is steeper than the same variation in the empirical probability. The values of pdf increase up to about $|DSN| = 1.3$ and then drop a little. Whether this is a real feature is uncertain. It is possible that it represents a spurious feature of the MEM as it tries to recover a uniform distribution at the edges, where there are fewer effective constraints.

For values $|DSN| > 60$ the MEM pdf diverges from the exponential approximations, which is supported by the empirical distribution of $DSN$. At the edge of the interval for $DSN$ considered for building the MEM pdf, a sharp increase appears. The latter is a spurious feature. In this area the conditions have very small values, such that the entropy exceeds the value of the quadratic error $E$ (see Appendix). Thus, the optimisation is done by maximising the entropy, which shifts the distribution toward a uniform distribution. The entropy may also affect the divergence from the exponential law for $|DSN|>60$; thus, we consider that the MEM pdf gives only a qualitative description for very high values of $|DSN|$. This is not very significant though, as the values of the pdf are below $10^{-4}$ in this region, far smaller than those close to $DSN = 0$.

\subsection{Time-scale Localisation of the Exponential Approximation}
\begin{figure}
 \centerline{ \includegraphics[width=1.0\textwidth,clip=]{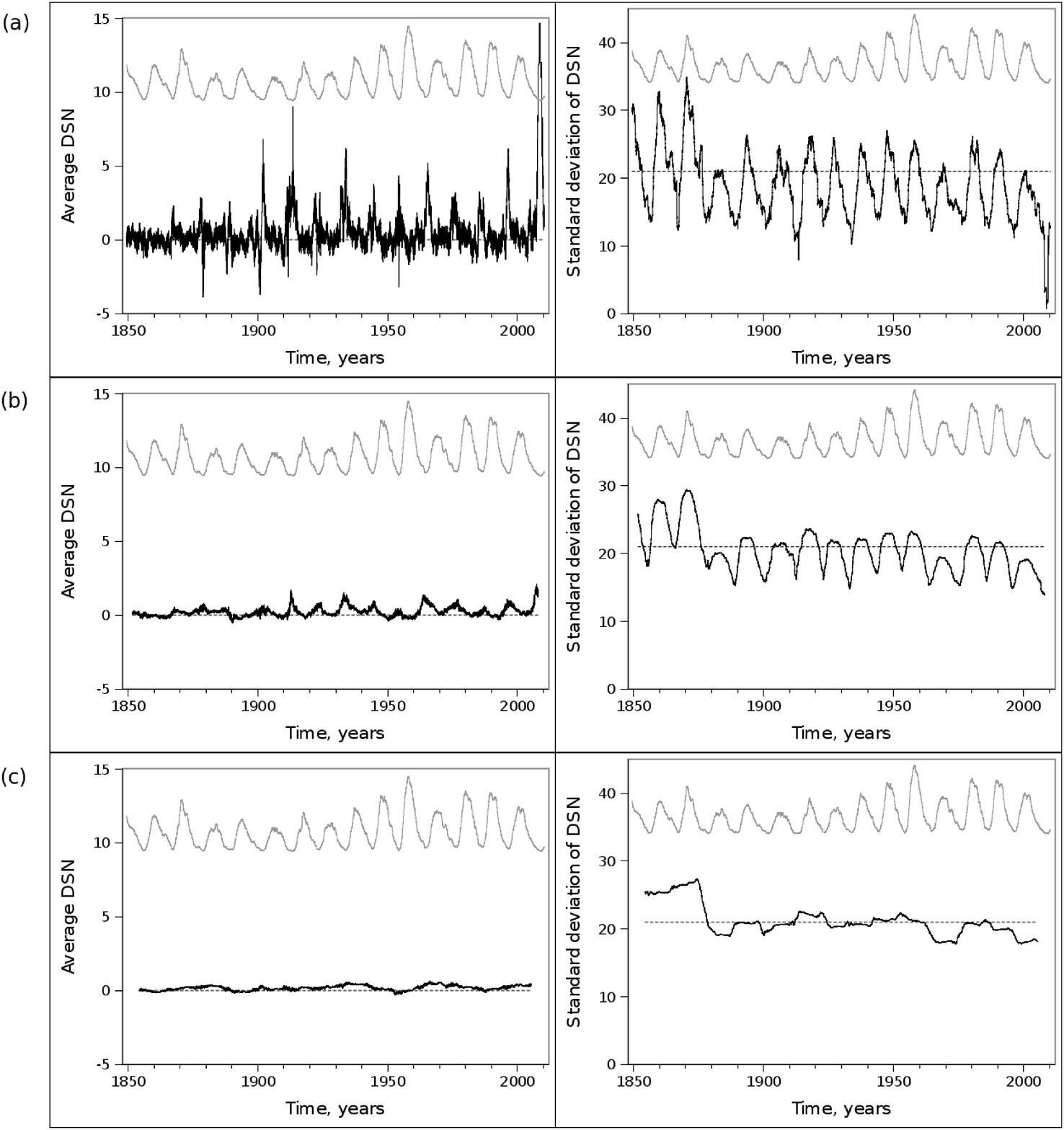} }
 \caption{Moving averages and standard deviations of $DSN$ (black lines) compared to the 1 year moving average $SN$ shape (grey line - not to scale). Only values $|DSN| \in [10;60]$ were considered in the computations. The statistical characteristics of $DSN$ were computed on time intervals of (a) 1 year, (b) 5.5 years and (c) 11 years. The dashed lines represent theoretical average and standard deviation computed with (\ref{Aprox}) and $m = \pm 9.35$, sign opposite to that of $DSN$. The average $SN$ describes the smoothed solar cycles' evolution.}
 \label{MedieSigma}
 \end{figure}

In order to determine the time scale at which the exponential law is valid, the moving average and standard deviation of $DSN$ were computed for different time intervals and their values were compared to theoretical values computed with (\ref{Exp}) and $m = \pm 9.35$, where the sign was taken $+$ for $DSN < 0$ and $-$ for $DSN > 0$. All computations were carried out for $|DSN| \in [10;60]$. The results are shown in Figure \ref{MedieSigma}. The best localisation, \textit{i.e.} the smallest time interval for which the distribution of $DSN$ is close to the theoretical distribution (\ref{Exp}), is obtained for a time span of about 11 years or approximately one solar cycle period. Cycles 10 and 11 stand out as anomalies, since their standard deviation is very different from the theoretical one even for large time intervals. For the Cycles 12-23, the standard deviation is close to the theoretical value with displacements of about 10\% from the theoretical standard deviation. The results obtained for a 1 year time interval show where deviations from the exponential law occur. The average $DSN$ has sharp peaks during cycles' minima. The standard deviation has maxima and minima corresponding to the maxima and minima of solar cycles. Its maxima are close to the theoretical standard deviation, while the minima are far from it. It appears thus that deviations from the exponential law occur around minima of the solar cycles. This is partly a consequence of the fact that the minima are populated with only a few values of $|DSN|$ above 10; thus, the sharp behaviour is partly generated by fluctuations in the average and standard deviation of $|DSN|$. 

\subsection{The Origin of the Exponential Law: Fluctuations vs. Deterministic SN}
\begin{figure} 
 \centerline{ \includegraphics[width=0.75\textwidth,clip=]{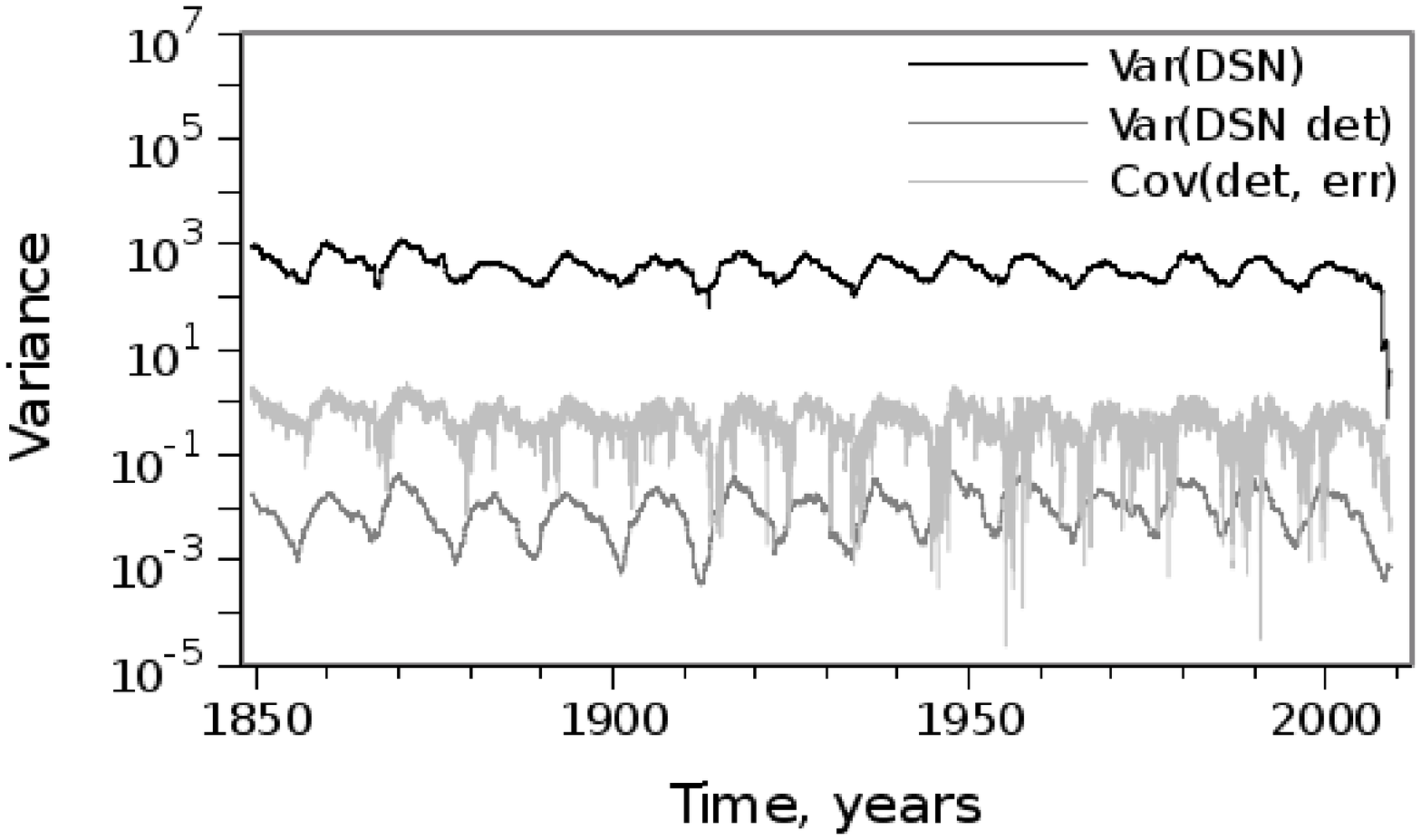} }
 \caption{The time evolution of the variance components of $DSN$: variance of total $DSN$, variance of deterministic $DSN$ (det) component as estimated from the one year average $SN$ and the covariance of the deterministic (det) and error (err) components of $DSN$. The difference between the total variance and the latter two components represents the variance of the error component (not represented), which was found to be superposed on the total variance. All values were computed on moving one year intervals.}
 \label{Varn}
 \end{figure}

The difference $DSN$ between $SN$ consecutive values at a fixed date can be reduced to two components: a smooth deterministic variation $DSN_{det}$ and an error (or fluctuation) term $DSN_{err}$, such that
\begin{equation}
DSN = DSN_{det} + DSN_{err}.
\end{equation}
A similar relation holds between the variances of the terms implied:
\begin{eqnarray}
Var(DSN) &=& Var(DSN_{det}) + Var(DSN_{err}) \nonumber \\
              && + Cov(DSN_{det},DSN_{err}),
\end{eqnarray}
where $Var(\cdot)$ represents the variance and $Cov(\cdot,\cdot)$ the covariance of the respective quantities. We tested to see whether one of the terms $DSN_{det}$, $DSN_{err}$ is far more important than the other. If this were the case, it is expected that the negligible term would have negligible statistical characteristics and its distribution would have a negligible contribution in the $DSN$ overall distribution. Since the averages of the $DSN$ components above are close to 0, the test was carried on their variances. Moving variances on intervals of one year were used. The results are shown in Figure \ref{Varn}. It can be seen that $Var(DSN)$ is at least 100 times greater than $Var(DSN_{det})$ or $Cov(DSN_{det},DSN_{err})$, thus by far the most important component of $DSN$ appears to be the fluctuation $DSN_{err}$. The exponential distribution can be assumed to be given by $DSN$ fluctuations and not its deterministic component.

\section{Conclusions}
The most important characteristic of the distribution of $DSN$ is its almost perfect symmetry with respect to the vertical axis. The small differences in the MEM pdf between negative and positive values of $DSN$ are not well supported by empirical values and we consider them spurious phenomena, due to the stochastic nature of the Simulated Annealing algorithm and to noise in the initial data.

Overall, the distribution of $DSN$ has several areas of specific variation: (i) For low values of $|DSN|$ the pdf decreases as $|DSN|$ increases, probably exponentially, but this is uncertain because of the small number of points. Around $|DSN| = 7$ the pdf has a slight but abrupt rise. (ii) For $|DSN|$ between 10 and 60 it obeys an exponential law of an approximate form:
\begin{equation} \label{Laplace}
\rho(DSN) = \frac{1}{Z} \exp \left( - \frac{|DSN|}{m} \right),
\end{equation}
with $m \approx 9.35$ experimentally and $m \approx 9.56$ in the MEM pdf; $Z$ is a constant. (iii) For very large $|DSN|$ the pdf decreases somewhat faster than exponential with increasing $|DSN|$.

The distribution (\ref{Laplace}) appears to be valid only when whole cycles are considered. For small time intervals, the $DSN$ distribution may deviate from the exponential law. This happens especially around minima of the solar cycles.

The presence of the exponential law dependence is interesting in itself. This result was previously found by \inlinecite{Lepreti} for $SN$ values spanning a 20 year period. Here the result is investigated in more detail and for a longer time period. The formula (\ref{Laplace}) represents a Laplace distribution centered at $DSN = 0$ \cite{Kotz}. The deviations from this law happen for very low values of $|DSN|$, which occur especially (but not only) in the minima of solar cycles, and also for very high values of $|DSN|$, which occur at the maxima of solar cycles. The latter case happens only for a few values of the solar cycles analysed, \textit{i.e.} for $|DSN| > 60$, as can be seen from Figure 1. Thus, it can be concluded that overall the exponential law holds for all values of $|DSN|$ of statistical significance, except values around $DSN = 0$. The failure of the exponential law for small $|DSN|$ may be due to the way the Sunspot Number ($SN$) is computed by counting both individual sunspots and groups of sunspots. This is especially significant for solar cycles minima, where abrupt jumps from $SN = 0$ to $SN \approx 7$ may easily appear. Nevertheless, the value $DSN = 0$ stays above the Laplace pdf (\ref{Laplace}).

The obtained results can be applied in the field of stochastic predictive models for solar activity and space weather. The daily stochastic behaviour is superposed over a long-term cyclic characteristic. These daily fluctuations may be represented as a stochastic process with jumps distributed according to (\ref{Laplace}). Nevertheless, such a stochastic process would give only an average, long-term representation. It must be completed with additional terms that give the exact distribution at smaller time scales. These terms are yet to be determined.

The Laplace distribution of $DSN$ is linked to the exponential distribution of $SN$, as evidenced by \inlinecite{Noble}. Indeed, by assuming that daily $SN$ values are independent and identically distributed random variables with exponential distributions, the difference of any two $SN$ variables is a random variable with a Laplace distribution. However, the $SN$ values are not independent on a day-to-day basis. Similarly, the Laplace distribution does not appear at a daily level, but rather at whole cycle levels. We hypothesize that the Laplace distribution of $DSN$ appears at time intervals for which the beginning and end values of $SN$ are completely independent. It is expected that the distribution is stationary at this time scale. This seems to be the case for 11 year intervals, as seen from Figure \ref{MedieSigma}, where both the average and standard deviation tend to become stable in time. Nevertheless, the standard deviation still has strong jumps even at this time scale, possibly showing abrupt changes in the Laplace distribution parameter $m$ from one cycle to the next. The long time span at which stability occurs may be a characteristic of the magnetic flux tubes lying in the convective layer, which is preserved at least until the magnetic field changes polarity. As they emerge through the photosphere, the flux tubes reveal this characteristic in the form of the Laplace distribution of $DSN$. Nevertheless, other more complex processes are at work for small time scales, where the distribution is highly irregular.

\appendix

\section{MEM Distributions}
The Maximum Entropy approach for probability distributions consists in determining the distribution that maximises the entropy with some imposed constraints. We present the way it was applied in the present work. Each pdf $\rho^-(DSN)$, $\rho^+(DSN)$ was determined separately as sets of probabilities $p^-(DSN_i)$, $p^+(DSN_j)$ associated with corresponding values $DSN_i<0$, $DSN_j>0$, $i,j = 1, 2, ..., N_P$, which were next transformed into pdf values. This is a general approach to building a continuous pdf which eliminates the need for a prior model to fit $\rho^-(\cdot)$, $\rho^+(\cdot)$. The number $N_P$ of values of $DSN$ considered in each case is much greater than the empirical number of probability values. The values $DSN_i$, $DSN_j$ where chosen at equal distances in the corresponding intervals.

We illustrate below the path followed for $DSN < 0$; for $DSN > 0$ we proceeded in a similar fashion. The Shannon discrete entropy was used:
\begin{equation}
H = - \sum_{i=1}^{N_P} p^-(DSN_i) \ln p^-(DSN_i).
\end{equation}
Next, some local constraints were imposed on the distribution. These were built with a Gaussian function:
\begin{equation}
f_k (x) = \exp \left( - \frac{(x - x^-_k)^2}{2 \sigma ^2} \right), k = 1,2,...,N_C.
\end{equation}
The points $x^-_k$ are equally distanced in the interval of $DSN<0$. The method of Kernel Density Estimation uses similar functions for building a pdf estimate with good results \cite{Rosenblatt,Parzen,Cranmer,Wu}. The functions $f_k$ were averaged, yielding empirical values $< f_k > _{emp}$ computed over $N^-$ empirical values of $x = DSN < 0$ collected from the 14 last cycles and also simulated averages $< f_k > _{sim}$ computed with the values $x = DSN_i$ and their associated probabilities $p^-(DSN_i)$:
\begin{eqnarray}
< f_k > _{emp} &=& \frac{1}{N^-} \sum _{x = \mathrm{empirical} \hspace{0.25mm} DSN} f_k (x), \\
< f_k > _{sim} &=& \frac{1}{N_P} \sum_{i=1}^{N_P} p^-(DSN_i) f_k (DSN_i).
\end{eqnarray}
A quadratic error was used:
\begin{equation}
E = \sum_{k=1}^{N_C} \left( < f_k > _{emp} - < f_k > _{sim} \right)^2.
\end{equation}
When the error E is minimised, the obtained distribution presents a lot of noise. In order to smooth it, the entropy H was maximised along with the minimisation of E. Thus, the function to be minimised was taken of the form:
\begin{equation}
E_H = E - k H,
\end{equation}
where $k$ is a constant that regulates the degree of smoothing. For big $k$ the distribution obtained is close to the uniform distribution, while for very small $k$ the distribution has a lot of noise. Based on tests carried on some known distributions, a value of $k = 10^{-6}$ was found to be sufficient.

In order to find the minimum of $E_H$ a modified version of the Simulated Annealing (SA) algorithm was used. In the original algorithm, a temperature parameter is decreased during the run in order to control the convergence towards the absolute minimum. We found this to be unnecessary, thus we set the temperature to 0. Also, in the original algorithm, at each step, the solution vector $(p^-(DSN_i))_i$ is constructed by modifying all its elements at random at each step. We found this to be detrimental and the modification was carried on only one element chosen at random at each step. The rate of change of the vector elements was decreased as the algorithm proceeded.

A number of $N_C = 101$ conditions were used in each case. An equal number of values of $x^-_k$, $x^+_k$ were generated in the intervals $[-100;-1]$ and $[1;120]$ respectively. In each case the solution was a vector of $N_P = 1\,000$ probabilities $\left( p^- (DSN_i) \right)_i$, $\left( p^+ (DSN_j) \right)_j$, corresponding to equally distanced values of $DSN$ set in the above intervals. Thus, in each interval there are about 10 times more simulated probability values for the pdf than empirical values. The conditions' parameter $\sigma$ was chosen such that each condition covers a significant part of the interval of values of $DSN$. It vas taken equal to $1/30$ of the length of the interval in each case. This gives a width at half-height for each $f_k$ of about 10 units of $DSN$ and a base coverage of about 20 units of $DSN$. The functions $f_k$ have maxima distanced at about 1 unit of $DSN$, thus yielding a dense set of conditions. This value of $\sigma$ was tested on some known distributions and was found satisfactory. \\

\textbf{Acknowledgements}
We are grateful to Professor Gelu M. Nita from the Center for Solar-Terrestrial Research of the New Jersey Institute of Technology for helpful discussions and suggestions regarding this work. We are also grateful to an unknown reviewer for advice that helped us significantly improve this paper.


\end{article} 


\begin{thebibliography}{}

\bibitem[\protect\citeauthoryear{{Aguirre, Letellier, and Maquet}}{2008}]{Aguirre} Aguirre, L.A., Letellier, C., Maquet, J.: 2008, Forecasting the Time Series of Sunspot Numbers, {\it Solar Phys.} \textbf{249}.
\bibitem[\protect\citeauthoryear{{Cerny}}{1985}]{Cerny} Cerny, V.: 1985, Thermodynamical approach to the traveling salesman problem: An efficient simulation algorithm, {\it Journal of Optimization Theory and Applications} \textbf{45}.
\bibitem[\protect\citeauthoryear{{Cranmer}}{2000}]{Cranmer} Cranmer, K. S.: 2000, Kernel Estimation in High-Energy Physics, {\it Computer Physics Communications} \textbf{136}.
\bibitem[\protect\citeauthoryear{{Greenkorn}}{2009}]{Greenkorn} Greenkorn, R.A.: 2009, Analysis of Sunspot Activity Cycles, {\it Solar Phys.} \textbf{255}.
\bibitem[\protect\citeauthoryear{{Ingber}}{1993}]{Ingber} Ingber, L.: 1993, Simulated annealing: Practice versus theory , {\it Mathematical and Computer Modelling} \textbf{18}.
\bibitem[\protect\citeauthoryear{{Jaynes}}{1957a}]{Jaynes1} Jaynes, E. T.: 1957a, Information Theory and Statistical Mechanics, {\it Phys. Rev.} \textbf{106}.
\bibitem[\protect\citeauthoryear{{Jaynes}}{1957b}]{Jaynes2} Jaynes, E. T.: 1957b, Information Theory and Statistical Mechanics II, {\it Phys. Rev.} \textbf{108}.
\bibitem[\protect\citeauthoryear{{Jaynes}}{1963}]{Jaynes} Jaynes, E. T.: 1963, Information Theory and Statistical Mechanics, in {\it Statistical Phys.}, K. Ford (ed.), Benjamin, New York.
\bibitem[\protect\citeauthoryear{{Kanazir and Wheatland}}{2010}]{Kanazir} Kanazir, M., Wheatland, M.S.: 2010, Time-Dependent Stochastic Modeling of Solar Active Region Energy, {\it Solar Phys.} \textbf{266}.
\bibitem[\protect\citeauthoryear{{Kirkpatrick, Gelatt, and Vecchi}}{1983}]{Kirk} Kirkpatrick, S., Gelatt, C. D., Vecchi, M. P.: 1983, Optimization by simulated annealing, {\it Science} \textbf{220}.
\bibitem[\protect\citeauthoryear{{Kotz, Kozubowski, and Podgorski}}{2001}]{Kotz} Kotz, S., Kozubowski, T.J., Podgorski, K.: 2001, The Laplace distribution and generalizations: a revisit with applications to Communications, Economics, Engineering and Finance, Birkh\"auser.
\bibitem[\protect\citeauthoryear{{Lepreti {\it et al.}}}{2000}]{Lepreti} Lepreti, F., Fanello, P.C., Zaccaro, F., Carbone, V.: 2000, Persistence of solar activity on small scales: Hurst analysis of time series coming from Hα flares, {\it Solar Phys.} \textbf{197}.
\bibitem[\protect\citeauthoryear{{Letellier {\it et al.}}}{2006}]{Letellier} Letellier, C., Aguirre, L.A., Maquet, J., Gilmore, R.: 2006, Evidence for low dimensional chaos in sunspot cycles, {\it Astron. Astrophys.} \textbf{449}.
\bibitem[\protect\citeauthoryear{{Noble and Wheatland}}{2011}]{Noble} Noble, P.L., Wheatland, M.S.: 2011, Modeling the Sunspot Number Distribution with a Fokker-Planck Equation, {\it Astrophys. J.} \textbf{732}.
\bibitem[\protect\citeauthoryear{{Nordemann}}{1992}]{Nordemann} Nordemann, D. J. R.: 1992, Sunspot number time series - Exponential fitting and solar behavior, {\it Solar Phys.} \textbf{141}.
\bibitem[\protect\citeauthoryear{{Parzen}}{1962}]{Parzen} Parzen, E.: 1962, On Estimation of a Probability Density Function and Mode, Ann. {\it Math. Statist.,} \textbf{33}. 
\bibitem[\protect\citeauthoryear{{Pontieri {\it et al.}}}{2003}]{Pontieri} Pontieri, A., Lepreti, F., Sorriso-Valvo, L., Vecchio, A., Carbone, V.: 2003, A Simple Model for the Solar Cycle, {\it Solar Phys.} \textbf{213}.
\bibitem[\protect\citeauthoryear{{Proctor and Gilbert}}{1994}]{Proctor} Proctor, M.R.E., Gilbert, A.D. (eds.): 1994, Lectures on Solar and Planetary Dynamos, Cambridge.
\bibitem[\protect\citeauthoryear{{Rosenblatt}}{1956}]{Rosenblatt} Rosenblatt, M.: 1956, Remarks on Some Nonparametric Estimates of a Density Function, {\it Ann. Math. Statist.} \textbf{27}.
\bibitem[\protect\citeauthoryear{{Salakhutdinova}}{1998}]{Salakhutdinova1} Salakhutdinova, I. I.: 1998, A Fractal Structure of the Time Series of Global Indices of Solar Activity, {\it Solar Phys.} \textbf{181}.
\bibitem[\protect\citeauthoryear{{Salakhutdinova}}{1999}]{Salakhutdinova2} Salakhutdinova, I. I.: 1999, Identifying the quasi-regular and stochastic components of solar cyclicity and their properties, {\it Solar Phys.} \textbf{188}.
\bibitem[\protect\citeauthoryear{{SIDC}}{2010}]{SIDC} SIDC-team, World Data Center for the Sunspot Index, Royal Observatory of Belgium, {\it Monthly Report on the International Sunspot Number}, online catalogue of the sunspot index: \url{http://www.sidc.be/sunspot-data/}, 1848-2010.
\bibitem[\protect\citeauthoryear{{Uffink}}{1995}]{Uffink} Uffink, J.: 1995, Can the Maximum Entropy Principle be explained as a consistency requirement?, {\it Studies in History and Philosophy of Modern Physics} \textbf{26B}.
\bibitem[\protect\citeauthoryear{{Vecchio {\it et al.}}}{2005}]{Vecchio} Vecchio, A., Primavera, L., Carbone, V., Sorriso-Valvo, L.: 2005, Periodic Behavior and Stochastic Fluctuations of Solar Activity: Proper Orthogonal Decomposition Analysis, {\it Solar Phys.} \textbf{229}.
\bibitem[\protect\citeauthoryear{{Wu and Mielniczuk}}{2002}]{Wu} Wu, W. B., Mielniczuk, J.: 2002, Kernel density estimation for linear processes, {\it Ann. Statist.} \textbf{30}. 


\end{thebibliography}
\end{document}